\documentstyle[12pt]{article}

\catcode`\@=11
\long\def\@makefntext#1{ 
\protect\noindent \hbox to 3.2pt {\hskip-.9pt
$^{{\ninerm\@thefnmark}}$\hfil}#1\hfill} 
\def\etal{et al.}
\def \eg {e.g.}
\def\ie{i.e.}
\def\thefootnote{\fnsymbol{footnote}}

\def\@makefnmark{\hbox to 0pt{$^{\@thefnmark}$\hss}}  

\def\ps@myheadings{\let\@mkboth\@gobbletwo
\def\@oddhead{\hbox{} 
\rightmark\hfil\ninerm\thepage}
\def\@oddfoot{}\def\@evenhead{\ninerm\thepage\hfil 
\leftmark\hbox{}}\def\@evenfoot{}
\def\sectionmark##1{}\def\subsectionmark##1{}}

\begin{document}

\newcommand{\symbolfootnote}{\renewcommand{\thefootnote}
	{\fnsymbol{footnote}}}
\renewcommand{\thefootnote}{\fnsymbol{footnote}}
\newcommand{\alphfootnote}
	{\setcounter{footnote}{0}
	 \renewcommand{\thefootnote}{\sevenrm\alph{footnote}}}

\newcounter{sectionc}\newcounter{subsectionc}\newcounter{subsubsectionc}
\renewcommand{\section}[1] {\vspace{0.6cm}\addtocounter{sectionc}{1}
\setcounter{subsectionc}{0}\setcounter{subsubsectionc}{0}\noindent
	{\bf\thesectionc. #1}\par\vspace{0.4cm}}
\renewcommand{\subsection}[1] {\vspace{0.6cm}\addtocounter{subsectionc}{1}
	\setcounter{subsubsectionc}{0}\noindent
	{\it\thesectionc.\thesubsectionc. #1}\par\vspace{0.4cm}}
\renewcommand{\subsubsection}[1]
{\vspace{0.6cm}\addtocounter{subsubsectionc}{1}
	\noindent {\rm\thesectionc.\thesubsectionc.\thesubsubsectionc.
	#1}\par\vspace{0.4cm}}
\newcommand{\nonumsection}[1] {\vspace{0.6cm}\noindent{\bf #1}
	\par\vspace{0.4cm}}

\newcounter{appendixc}
\newcounter{subappendixc}[appendixc]
\newcounter{subsubappendixc}[subappendixc]
\renewcommand{\thesubappendixc}{\Alph{appendixc}.\arabic{subappendixc}}
\renewcommand{\thesubsubappendixc}
	{\Alph{appendixc}.\arabic{subappendixc}.\arabic{subsubappendixc}}

\renewcommand{\appendix}[1] {\vspace{0.6cm}
        \refstepcounter{appendixc}
        \setcounter{figure}{0}
        \setcounter{table}{0}
        \setcounter{equation}{0}
        \renewcommand{\thefigure}{\Alph{appendixc}.\arabic{figure}}
        \renewcommand{\thetable}{\Alph{appendixc}.\arabic{table}}
        \renewcommand{\theappendixc}{\Alph{appendixc}}
        \renewcommand{\theequation}{\Alph{appendixc}.\arabic{equation}}
        \noindent{\bf Appendix \theappendixc #1}\par\vspace{0.4cm}}
\newcommand{\subappendix}[1] {\vspace{0.6cm}
        \refstepcounter{subappendixc}
        \noindent{\bf Appendix \thesubappendixc. #1}\par\vspace{0.4cm}}
\newcommand{\subsubappendix}[1] {\vspace{0.6cm}
        \refstepcounter{subsubappendixc}
        \noindent{\it Appendix \thesubsubappendixc. #1}
	\par\vspace{0.4cm}}

\def\abstracts#1{{
	\centering{\begin{minipage}{30pc}\tenrm\baselineskip=12pt\noindent
	\centerline{\tenrm ABSTRACT}\vspace{0.3cm}
	\parindent=0pt #1
	\end{minipage} }\par}}

\renewenvironment{thebibliography}[1]
	{\begin{list}{\arabic{enumi}.}
	{\usecounter{enumi}\setlength{\parsep}{0pt}
\setlength{\leftmargin 1.25cm}{\rightmargin 0pt}
	 \setlength{\itemsep}{0pt} \settowidth
	{\labelwidth}{#1.}\sloppy}}{\end{list}}

\topsep=0in\parsep=0in\itemsep=0in
\parindent=1.5pc

\newcounter{itemlistc}
\newcounter{romanlistc}
\newcounter{alphlistc}
\newcounter{arabiclistc}
\newenvironment{itemlist}
    	{\setcounter{itemlistc}{0}
	 \begin{list}{$\bullet$}
	{\usecounter{itemlistc}
	 \setlength{\parsep}{0pt}
	 \setlength{\itemsep}{0pt}}}{\end{list}}

\newenvironment{romanlist}
	{\setcounter{romanlistc}{0}
	 \begin{list}{$($\roman{romanlistc}$)$}
	{\usecounter{romanlistc}
	 \setlength{\parsep}{0pt}
	 \setlength{\itemsep}{0pt}}}{\end{list}}

\newenvironment{alphlist}
	{\setcounter{alphlistc}{0}
	 \begin{list}{$($\alph{alphlistc}$)$}
	{\usecounter{alphlistc}
	 \setlength{\parsep}{0pt}
	 \setlength{\itemsep}{0pt}}}{\end{list}}

\newenvironment{arabiclist}
	{\setcounter{arabiclistc}{0}
	 \begin{list}{\arabic{arabiclistc}}
	{\usecounter{arabiclistc}
	 \setlength{\parsep}{0pt}
	 \setlength{\itemsep}{0pt}}}{\end{list}}

\newcommand{\fcaption}[1]{
        \refstepcounter{figure}
        \setbox\@tempboxa = \hbox{\tenrm Fig.~\thefigure. #1}
        \ifdim \wd\@tempboxa > 6in
           {\begin{center}
        \parbox{6in}{\tenrm\baselineskip=12pt Fig.~\thefigure. #1 }
            \end{center}}
        \else
             {\begin{center}
             {\tenrm Fig.~\thefigure. #1}
              \end{center}}
        \fi}

\newcommand{\tcaption}[1]{
        \refstepcounter{table}
        \setbox\@tempboxa = \hbox{\tenrm Table~\thetable. #1}
        \ifdim \wd\@tempboxa > 6in
           {\begin{center}
        \parbox{6in}{\tenrm\baselineskip=12pt Table~\thetable. #1 }
            \end{center}}
        \else
             {\begin{center}
             {\tenrm Table~\thetable. #1}
              \end{center}}
        \fi}

\def\@citex[#1]#2{\if@filesw\immediate\write\@auxout
	{\string\citation{#2}}\fi
\def\@citea{}\@cite{\@for\@citeb:=#2\do
	{\@citea\def\@citea{,}\@ifundefined
	{b@\@citeb}{{\bf ?}\@warning
	{Citation `\@citeb' on page \thepage \space undefined}}
	{\csname b@\@citeb\endcsname}}}{#1}}

\newif\if@cghi
\def\cite{\@cghitrue\@ifnextchar [{\@tempswatrue
	\@citex}{\@tempswafalse\@citex[]}}
\def\citelow{\@cghifalse\@ifnextchar [{\@tempswatrue
	\@citex}{\@tempswafalse\@citex[]}}
\def\@cite#1#2{{$\null^{#1}$\if@tempswa\typeout
	{IJCGA warning: optional citation argument
	ignored: `#2'} \fi}}
\newcommand{\citeup}{\cite}

\def\fnm#1{$^{\mbox{\scriptsize #1}}$}
\def\fnt#1#2{\footnotetext{\kern-.3em
	{$^{\mbox{\sevenrm #1}}$}{#2}}}

\font\twelvebf=cmbx10 scaled\magstep 1
\font\twelverm=cmr10 scaled\magstep 1
\font\twelveit=cmti10 scaled\magstep 1
\font\elevenbfit=cmbxti10 scaled\magstephalf
\font\elevenbf=cmbx10 scaled\magstephalf
\font\elevenrm=cmr10 scaled\magstephalf
\font\elevenit=cmti10 scaled\magstephalf
\font\bfit=cmbxti10
\font\tenbf=cmbx10
\font\tenrm=cmr10
\font\tenit=cmti10
\font\ninebf=cmbx9
\font\ninerm=cmr9
\font\nineit=cmti9
\font\eightbf=cmbx8
\font\eightrm=cmr8
\font\eightit=cmti8


\centerline{\tenbf ROTATION CURVES OF 967 SPIRAL GALAXIES:  }
\baselineskip=22pt
\centerline{\tenbf IMPLICATIONS FOR DARK MATTER}
\baselineskip=16pt
\vspace{0.8cm}
\centerline{\tenrm MASSIMO PERSIC, PAOLO SALUCCI \& FULVIO STEL}
\baselineskip=13pt
\centerline{\tenit SISSA, Via Beirut 4, 34013 Trieste, Italy}
\vspace{0.9cm}

\abstracts{We present the rotation curves of 967 spiral galaxies, obtained by
deprojecting
and folding the raw H$\alpha$ data  published by Mathewson \etal
(1992). Of these, 80 meet objective excellence criteria and are suitable for
individual detailed mass modelling, while 820 are
suitable for statistical studies. A
preliminary analysis of their properties confirms that rotation curves
are a universal
function of luminosity and that the dark matter fraction in spirals increases
with decreasing luminosity.}

\section{ Introduction}

Rotation curves (hereafter RCs) are the prime mass tracers within spiral
galaxies. Therefore, knowledge of their morphology and structural implications
is essential for theories/experiments concerning
galaxy formation (\eg: Cen \& Ostriker
1993; Navarro \& White 1994; Evrard \etal 1994). Such a knowledge is of course
crucially increased when large samples of good-quality curves become available.

The H$\alpha$ velocities referred to the plane of the sky,
 of nearly one thousand spirals published by
Mathewson \etal (1992; hereafter MFB) represent by far the largest available
sample of
(raw) measurements of galaxy rotation. However, it is
well known that  recessional velocities, which are adequate for the purpose
of estimating
the maximum circular velocity, require a careful
treatment before they can yield the actual  rotation curves.
 In a related paper (Persic \& Salucci 1995;
hereafter PS95),  after folding, deprojecting
and smoothing the raw MFB data we work out the actual RC's.
 Because of its size, homogeneity, quality, and
spanned range of luminosities and asymptotic velocities, the sample of RCs thus
obtained will serve as a main database for studies of galaxy structure.

The plan of this contribution is as follows. In section 2 we outline the
procedure used to obtain the RCs from the raw data. In section 3 we classify
the 967 RCs into three quality subsets. Section 4 briefly illustrates the
results of a preliminary analysis of these RCs.
 The 967 rotation curves
will be  published in Persic and Salucci, 1995 (PS95).

\section{ Data Analysis}

For details of the observations and data acquisition and reduction, the reader
is
referred to MFB. Here we outline the basic procedure leading to the
final RCs. For each RC it includes: 1) a selection of the individual velocity
data; and 2) the identification of the kinematical center about which to fold
such data. In addition, step 3) in order to evaluate the
extension of each RC, we have compared the radius corresponding to the
farthest measured velocity with the optical size $R_{opt}$.
\begin{figure}
\vspace{7cm}
\includegraphics{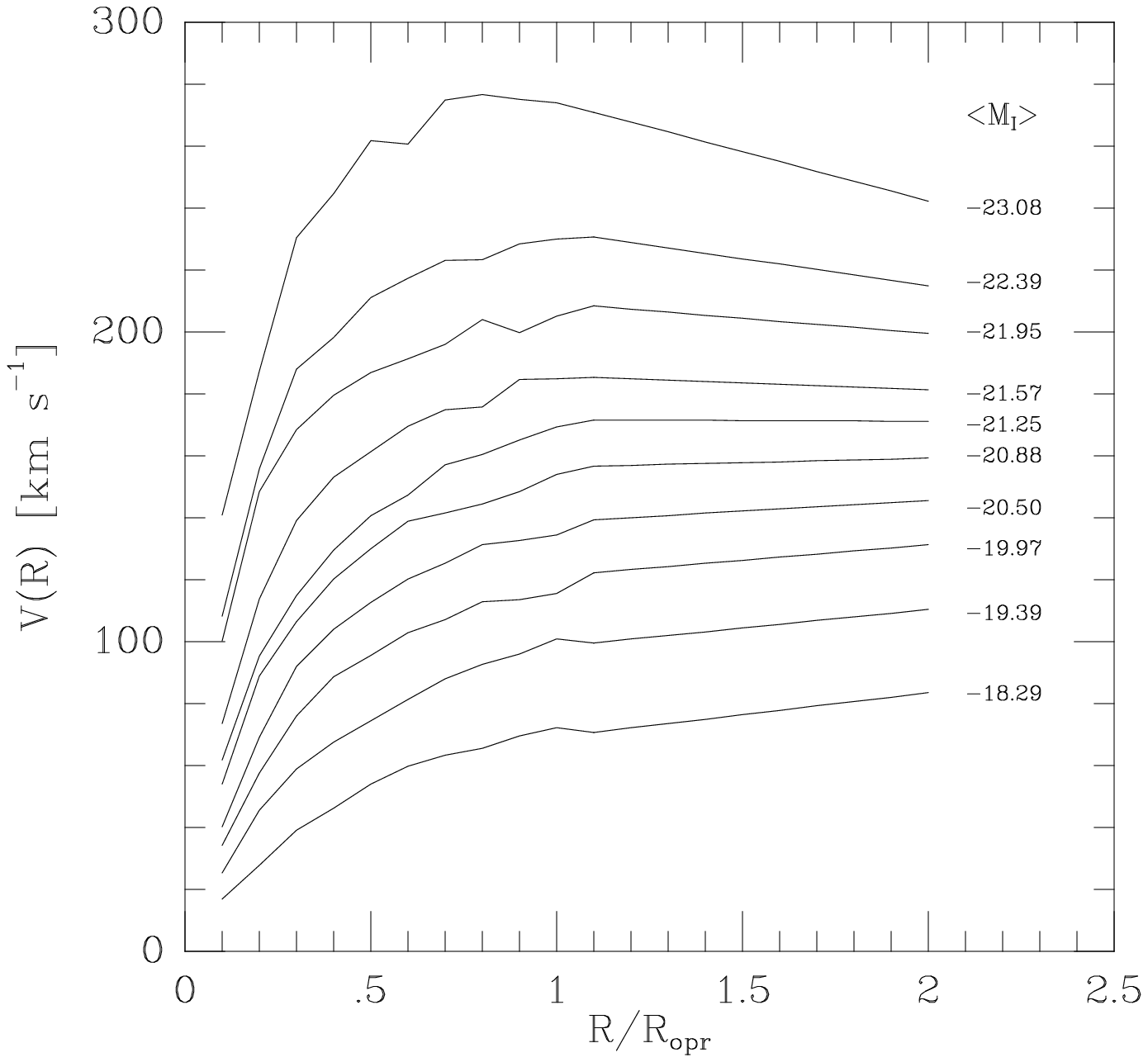}
\caption{Sinthetic rotation curves resulting from the coaddition pf the 616
RC's from PS95. Galactocentric radii are normalized to the optical radius }
\label{fig:match}
\smallskip
\medskip
\end{figure}
1) The quality of each velocity point is measured by the cross-correlation
coefficient $\rho$ (given in the public MFB data files) between the observed
and  template H$\alpha$ line profiles. Based on the rms velocity error as a
function of $\rho$, we have found that a good correlation ($\rho > 0.35$) is
essential in order to ensure a high reliability of the measure.

2) The photometric center (the point of maximum emission in the $I$-band) does
not necessarily coincide with the dynamical center of the galaxy. To
obtain the kinematical center we have proceeded as follows: we have assumed
that, once folded, each curve must be perfectly symmetric around its
kinematical center, so that the curve defined by the approaching side coincides
(within errors) with the one traced by the receding side. In practice, we have
started from the photometric center; then, if warranted, we have estimated
the kinematic center by folding the data around slightly different zero-points
until the global symmetry of the velocity arms is maximized. Table 1 of PS95
contains the solution of this folding procedure, \ie: the heliocentric systemic
velocity, and the offset of the kinematic center from the photometric center.

3) It is essential to have a reference scale for the optical size of each
galaxy in order to assess the extension of the RC as compared with the visible
 matter. For
this purpose, using the {\it I}--band CCD surface photometry of MFB, we have
computed the radius, $R_{opt}$, encompassing 83\% of the integrated light (see
Table 1 of PS95; for an exponential disk this corresponds to 3.2 lengthscales,
which in turn corresponds, for a Freeman disk, to the de Vaucouleurs 25
$B$-mag/arcsec$^2$ photometric radius.)

\section{ The Rotation Curves}

We have identified 900 RCs that are symmetric, with reasonably low rms internal
scatter and reasonable high sampling density. These curves are shown in Fig.1
of PS95. This set of 900 safe RCs can be split into a subset of 80 {\it
excellent} curves (Set A), and one of 820 {\it fair} curves (Set B).
The 80 RCs in Set A have the following properties, which ensure that they are
good tracers of the gravitational potential in the region of interest and hence
are well suited for  detailed mass modelling: 1) the approaching
and receding sides are very symmetric; 2) the data are extended out to (at
least) $R_{opt}$; and 3) there are $\geq 30$ data points homogeneously
distributed with radius and per arm. For each galaxy, from the folded RC we
produce a smooth RC as follows. We smooth the folded velocities by binning the
$\leq N$ nearest data points contained within a fixed maximum bin size $W$. The
values used for $W$ are mostly $0.050\,R_{opt}$ and $0.075\,R_{opt}$, and for
$N$ are mostly 4 and 6. These values are chosen for each RC according to its
sampling density: however, the profiles of the RCs are not significantly
changed with other
(reasonable) choices as can be seen by comparing the smoothed RCs with the
original folded ones. In each bin we compute the average rotation velocity and
its uncertainty. The resulting smoothed curves are shown in Fig.2 of PS95.
\begin{figure}
\vspace{13cm}
\includegraphics{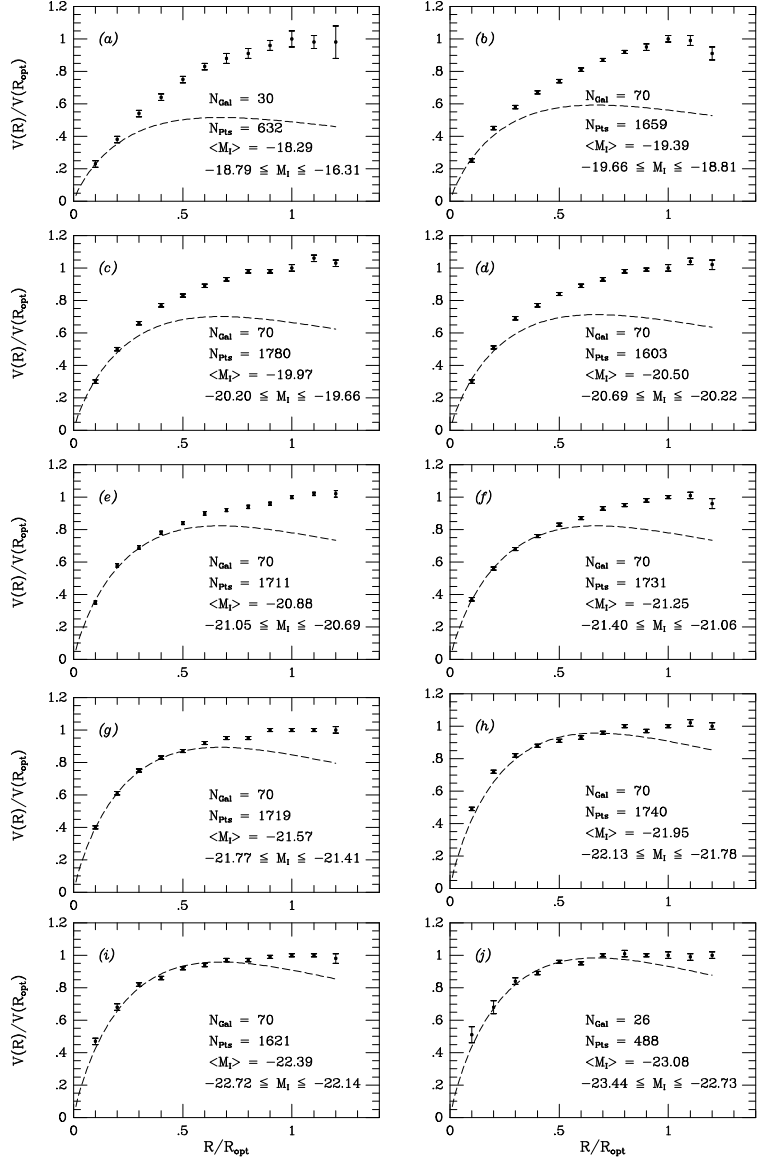}

\caption{Maximum disk mass-modelling (dashed lines) of the synthetic rotation
curves}
\medskip
\label{fig:match}
\end{figure}
The 820 RCs in Set B fail, to various extents, at least one of the criteria
stated above and therefore may be not suitable for accurate and direct mass
modelling. However, they constitute a large database for those methods able to
recover the DM properties with less stringent requirements on the RC
quality.

Finally, the 67 RCs of Set C have severe global asymmetries, large rms internal
scatter, insufficient sampling and/or large-scale deviations from circular
motion. These curves may be useful for studing the non-axisymmetric
disturbances
(\eg, bars, spiral arms) in spiral galaxies and are shown in Fig.3 of PS95.

\section { Discussion.}

For size, homogeneity, and intrinsic quality of the individual curves, the
sample of 967 RCs of PS95 constitutes by far  the best sample of RCs available
to date. As such, it offers a unique opportunity for investigating in
considerable depth the properties of dark matter in galaxies, \eg  its radial
distribution, the total quantity, and the scaling laws of the structural
parameters (see review
by Ashman 1992). As  part of a preliminary study (Persic, Salucci \& Stel
1995), we have co-added the 616 RCs (out of the 900) whose extension
exceeds $0.8 \,R_{opt}$, to form 10 synthetic curves (see Rubin \etal  1985),
each
related to a different range of luminosity (spanning approximately 5 magnitudes
overall) and extending out to $1.6 \,R_{opt}$ (see Fig.1).

We find that the RCs are
well represented by a double straight line, with the change of slope occurring
at about $R_{opt}$ (see Fig.1). In detail, RCs depend on radius and galaxy
luminosity according to:
\begin{equation}
V(r) ~ = ~ V(R_{opt}) \biggl[1 +  \biggl(0.30 - 0.27 \,
	   {\rm log}\,{L_I \over 10^{10}L_{\odot}}\biggr)
	   \biggl( {r \over R_{opt}} -1 \biggr) \biggr]
		~~.~.~.~ 0.4 \leq r/R_{opt} \leq 1.0
\end{equation}
\begin{equation}
V(r) ~ = ~ V(R_{opt}) \biggl[1 +  \biggl(0.12 - 0.16 \,
	   {\rm log}\,{L_I \over 10^{10}L_{\odot}}\biggr)
	   \biggl( {r \over R_{opt}} -1 \biggr)
           \biggr]
		~~.~.~.~ 1.0 \leq r/R_{opt} \leq 1.6
\end{equation}
where $V(R_{opt})=100 ~(L_I/10^{10}L_{\odot})^{0.34}$ km s$^{-1}$.
 We definitely confirm the strong luminosity
dependence of the RC shape claimed by Rubin \etal (1985) and supported by
Persic \& Salucci (1991) and Casertano \& van Gorkom (1991).

We have tried to reproduce the rotation curves with only the luminous matter,
assumed as an exponential thin disk. In Figure (2) we plot  the contributions
of the visible matter
(dash lines) alongside the coadded rotation curves
of spirals of differnt luminosity. We note that  at about $R_{opt}$ the former
is always unable to reproduce
the observed kinematics. Moreover, while in high-luminosity
galaxies, the stellar disk reproduces  the RC inside $r \simeq {2
\over 3 }
R_{opt}$ extremely well, in low-luminosity ones it is unable to do so at
any radius.  Thus, we  find that the dark component is
ubiquitous in spirals and that the dark matter fraction inside $R_{opt} $
increases with
decreasing luminosity as  ${V^2_{disk}\over V^2}|_{R_{opt}} \simeq 0.2+0.13 (M_
+23$, in agreement with
previous results (\eg: Persic \& Salucci 1988, 1990; Salucci \& Frenk
1989; Salucci \etal 1991; Casertano \& van Gorkom 1991; Broeils 1992).
(For a more detailed mass modelling of spirals see Persic, Salucci, Stel,
1996).
\begin{figure}
\vspace{8cm}
\includegraphics{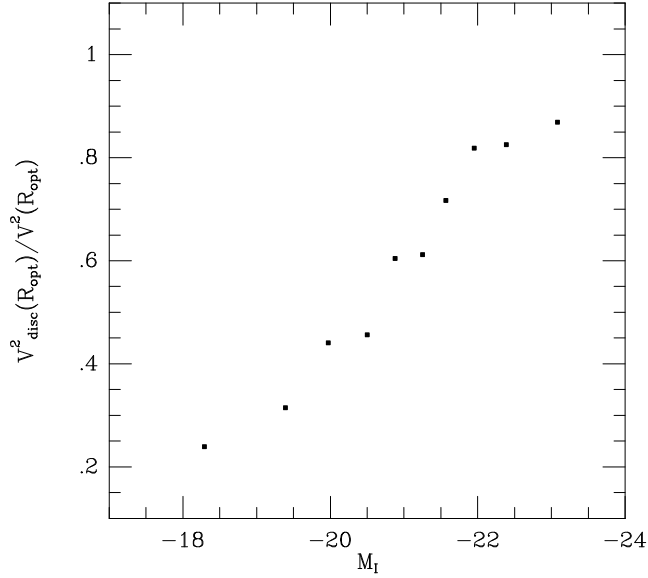}

\caption{Disk-to-total velocity (mass) ratio as a function of I-Magnitude}
\label{fig:match}
\medskip
\smallskip
\end{figure}
 In order to study the DM distribution we have calculated the virial
 radius $R_V$ of the dark halos associated to
the various RCs, and have rescaled each curve according to its $R_V$.
Avereging the halo overdensity with a top-hat window
$<\rho_H>_{{4\over 3}\pi R_V^3}\,=\,200\cdot\rho_c$ (where $\rho_c$ is the
critical density), we obtain $R_V$ (see Fall  \& Efstathiou 1980)
\begin{equation}
{3\over 4\pi G R_V}\Bigl[\,{V^2(R_V)\over R_V}\,-\, {M_D\over R_D^2} G x
 \bigl[I_0K_0-I_1K_1\bigr]_x \,\Bigr]\,= 200 \rho_c
\end{equation}
where $R_D$ and $M_D$ are respectively the  scale--length and
the total mass of the galactic disks, while $I_0$, $I_1$, $K_0$ and  $K_1$ are
the modified Bessel functions calculated at $x\,=\,{R_V \over 2 R_D}$. In this
way we can see (Fig. 5) that  RCs
associated with low luminosity spirals are similar
to the inner parts of RCs  associated with  higt luminosity spirals.

\begin{figure}
\newpage
\vspace{7cm}
\includegraphics{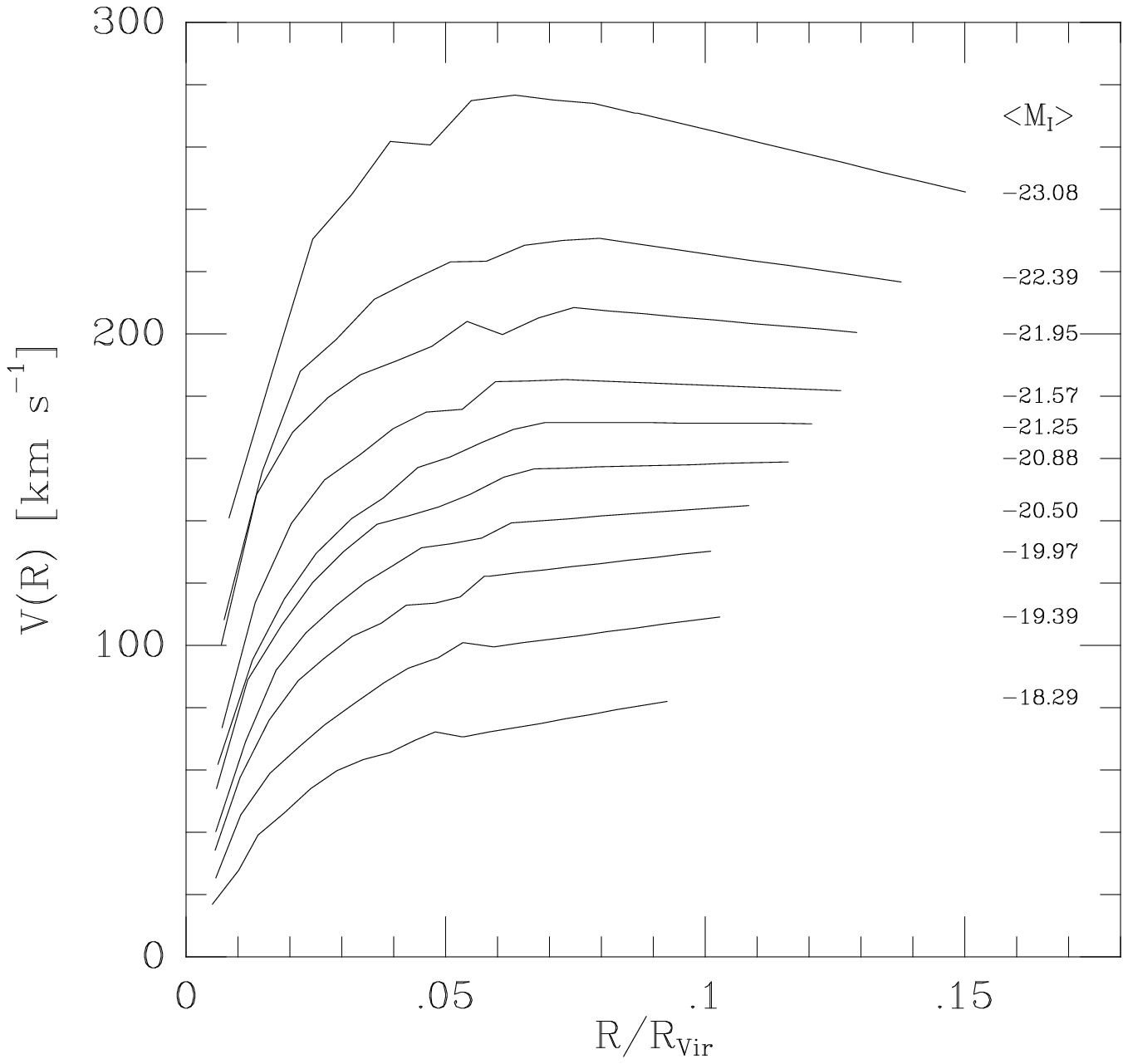}
\addtocounter{figure}{1}
\caption{Sinthetic rotation curves resulting from the coaddition of the 616
RC's from PS95. Galactocentric radii are normalized to the virial radius}
\medskip
\label{fig:match}
\end{figure}
\noindent This  evidence could
be interpreted as a consequence of the self--similarity of dark halos
in spiral galaxies.  The {\it differences in shape} of RC, which are
evident when  we frame them on the visible matter length-scale depends on
 the fact that doing so we compare {\it different parts}
 of dark halos. In fact, while dark halos are self-similar,
    their interplay
 with the visible matter is luminosity-dependent.
Finally, in  Goldshmith et al, we have compared  Cold Dark Matter
simulated halos with the above  RC's (see  Fig. 4) and have demonstrated
that CDM
halos  reach  their asymptotical velocity too quickly. This is particularly
evident in low luminosity RCs which beeing halo dominated, should better
 reproduce the halo RC.
\begin{figure}
\vspace{13cm}
\includegraphics{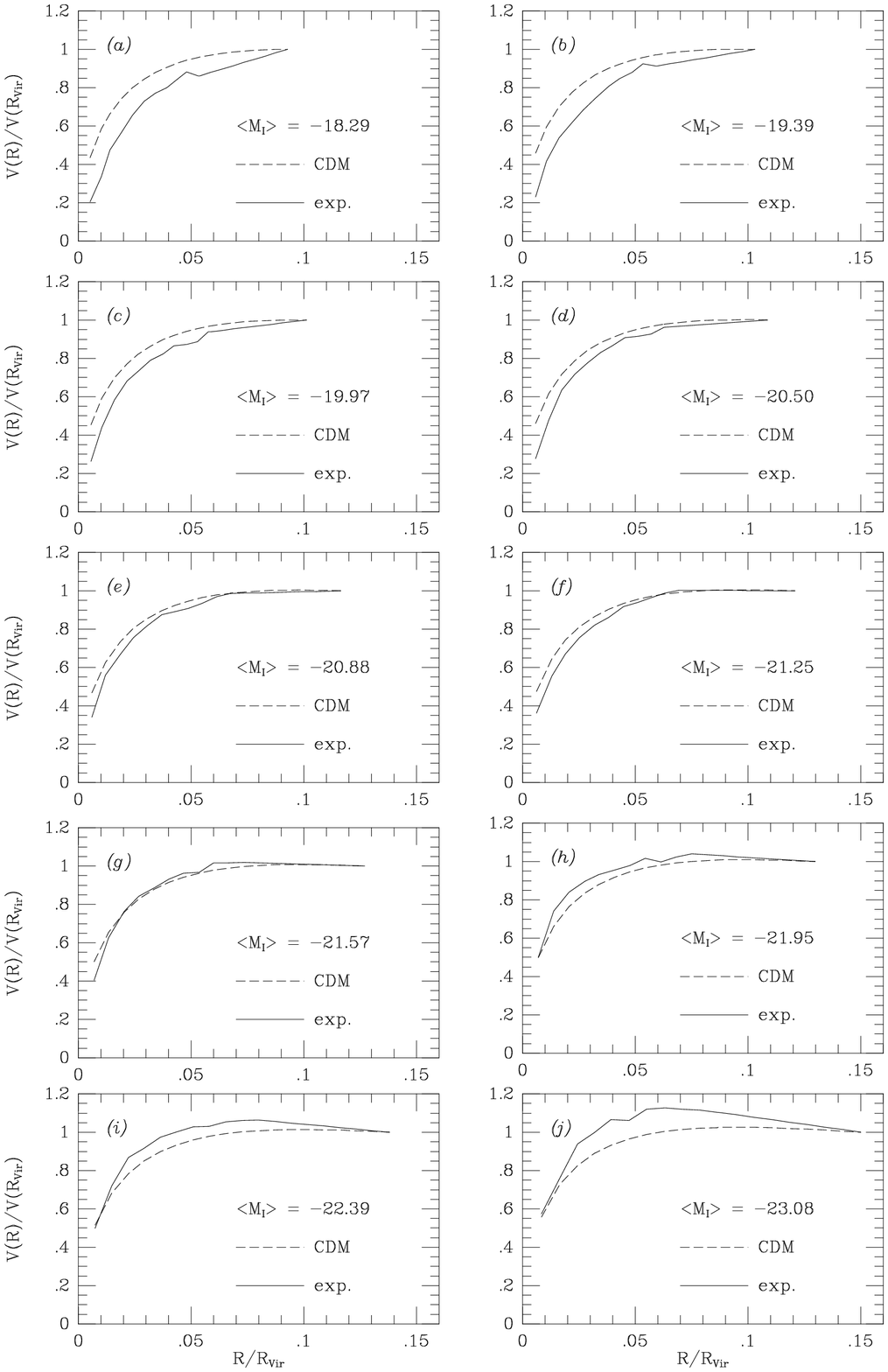}
\addtocounter{figure}{-2}
\caption{ Comparison between the synthetic rotation curves and CDM predictions
  }
\label{fig:match}
\smallskip
\medskip
\end{figure}

\end{document}